\documentstyle[12pt,titlepage]{article}
\begin{document}
\pagestyle{empty}
\newcommand{\hf}{\hfill}
\newcommand{\lapproxeq}{\lower
.7ex\hbox{$\;\stackrel{\textstyle <}{\sim}\;$}}
\baselineskip=0.212in

\begin{flushleft}
\large
{SAGA-HE-68-94
   \hfill September, 1994}  \\
%%% \hfill \today}  \\
\end{flushleft}

\vspace{2.5cm}

\begin{center}

\LARGE{{\bf Parton Distributions in Nuclei}} \\

\vspace{2.0cm}

\Large
{S. Kumano $^\star$ }         \\

\vspace{1.3cm}

\Large
{Department of Physics}         \\

\vspace{0.1cm}

\Large
{Saga University}      \\

\vspace{0.1cm}

\Large
{Saga 840, Japan} \\

\vspace{2.8cm}

\large
{Plenary Talk given} \\

\vspace{0.3cm}

{at the International Symposium on Medium Energy Physics} \\

\vspace{0.7cm}

{Beijing, P. R. China,  August 22--26, 1994}  \\

\end{center}

\vspace{2.5cm}
\vfill

{\rule{6.cm}{0.1mm}} \\

\vspace{-0.4cm}

\normalsize
{$\star$ Email: kumanos@himiko.cc.saga-u.ac.jp}  \\

\vspace{-0.2cm}
\hfill
{to be published in proceedings}

\vfill\eject
%%%%%%%%%%%%%%%%%%%%%%%%%%%%%%%%%%%%%%%%%%%%%%%%%%%%%%%%%%%%%%%%%%%%%%
%%%%%%%%%%%%%%%%%%%%%%%%%%%%%%%%%%%%%%%%%%%%%%%%%%%%%%%%%%%%%%%%%%%%%%
\pagestyle{plain}
\begin{center}

\Large
{Parton Distributions in Nuclei} \\

\vspace{0.5cm}

{S. Kumano $^*$}             \\

{Department of Physics}    \\

{Saga University}      \\

{Saga 840, Japan} \\

\vspace{0.7cm}

\normalsize
Abstract
\end{center}
\vspace{-0.46cm}

We discuss two topics in this talk. One is a brief overview
of nuclear parton distributions, and the other is
SU(2)-flavor-symmetry breaking in {\it nuclear} antiquark
distributions.
First, we show that nuclear structure functions $F_2^A(x)$
could be explained in a $Q^2$ rescaling model with
parton recombination effects.
Then, nuclear gluon distributions are discussed
in this parton model.
We point out other interesting distributions
for future investigations.
Second, $\bar u-\bar d$ distributions in nuclei
are discussed in order to shed light on nuclear modification
of the $\bar u-\bar d$ distribution.

\vspace{0.8cm}

\noindent
{\bf 1. Overview of nuclear parton distributions}

\vspace{0.2cm}

Modifications of the structure function $F_2(x)$
in nuclei were discovered by
European Muon Collaboration (EMC effect).
This effect has been an interesting topic
because it may provide an explicit quark signature
in nuclear phenomena.
Most of these investigations discuss a
``global'' EMC effect in the sense that the effect is
averaged over all constituents in a nucleus.
However, it is interesting to investigate possible
semi-inclusive or semi-exclusive processes
for finding a ``local'' EMC effect [1]
and possible relations between a local gluonic
EMC effect and the $J/\psi$ suppression.

In recent years, many experimental data of $F_2^A(x)/F_2^D(x)$
are obtained in the small $x$ region,
which is so called the shadowing region.
There are theoretical attempts to explain the EMC effect
in the medium-large $x$ region and the shadowing in the small $x$.
However, most models investigated both regions separately, and
enough effort has not been made for studying the structure function
in the whole $x$ region.
It is not straightforward to study a model which is valid
in the wide $x$ region.
An attempt to combine the medium-$x$ physics and the small-$x$ one
in a dynamically consistent way is made in Ref. [2].
Our model is based on a parton model and is not on macroscopic
nuclear physics (nuclear binding, Fermi motion, vector-meson
dominance, and so on) explicitly.
We simply incorporated two mechanism in our model:
Q$^2$ rescaling and parton recombination.
The rescaling has been used for explaining the EMC effect
in the medium $x$ region and the recombination for
the shadowing in the small $x$.
We calculated nuclear parton distributions
by using these ideas at small Q$^2$ and obtained distributions
are evolved by using the Altarelli-Parisi equation.

\vspace{-0.46cm}
%%%\begin{list}{}{\leftmargin 0.0in \rightmargin 3.7in}
\begin{list}{}{\leftmargin 0.0in \rightmargin 3.3in}
\item
$~~~~$
As a result, we obtained reasonably
good agreement with the experimental
data in the region ($0.005<x<0.8$) as shown in Fig. 1.
In the large $x$ region, the ratio ($F_2^A(x)/F_2^D(x)>1$)
is explained by quark-gluon recombinations, which produce
results similar to those by the nucleon Fermi motion.
In the medium $x$ region, the EMC effect is mainly due to
the $Q^2$ rescaling mechanism in our model. In the small $x$ region,
shadowing effects are obtained by the parton recombinations.
However, our shadowing at very small $x(<0.02)$ is very sensitive
to the input gluon distribution.
\end{list}
\vspace{-1.0cm}
\vspace{-1.5cm}
\hspace{7.8cm}
Fig. 1 Comparison with $F_2^{Ca} /F_2^D$ data [2].

\vspace{+1.5cm}
\vspace{-0.46cm}
\begin{list}{}{\leftmargin 0.0in \rightmargin 3.3in}
\item
$~~~~$
Q$^2$ variations of the nuclear structure function
ratio $F_2^{Ca}(x,Q^2)/F_2^D(x,Q^2)$ are calculated
in our parton model.
Calculated results are compared with
the New Muon Collaboration (NMC) data in Fig. 2.
Our theoretical results show small Q$^2$ variation and
are consistent with existing experimental data [2].
The detailed analysis of the $Q^2$ evolution
is still in progress [3].
\end{list}
\vspace{-0.3cm}
\vspace{-1.0cm}
\hspace{8.5cm}
Fig. 2 Q$^2$ variation of $F_2^{Ca}/F_2^D$ [2].
\vspace{+0.3cm}

We find that the model can explain
experimental $F_2^A(x)$ structure functions
in the wide Bjorken$-x$ range ($0.005<x<0.8$).
The structure function $F_2(x)$ has been well investigated
both experimentally and theoretically.
However, there are nuclear structure functions,
which are little understood.
We summarize the current status of nuclear parton distributions
in Table 1
in the Michelin style.

Because $F_2(x)$ in the medium $x$ is dominated by
valence-quark distributions [$q_v(x)$], we consider that
$q_v(x)$ at medium $x$ is well known.
On the other hand, $q_v(x)$ at small $x$ is not understood.
We don't know whether valence-quark shadowing is similar
to the $F_2$ shadowing.
It could be very important to study the valence-quark shadowing
by measuring $F_3^A/F_3^D$ in neutrino experiments.

Sea-quark distributions [$q_s(x)$] are dominant in $F_2(x)$ at small $x$,
so that we know the behavior of $q_s(x)$ at small $x$.
In the region of $x=0.1-0.2$, we have Fermilab-E772 Drell-Yan data
which indicates small nuclear modifications of the sea distribution
in the iron nucleus. However, the situation is not very clear
if we look at the Drell-Yan data in the carbon and calcium nuclei.
We need accurate data especially the data which show
the A dependence.
Furthermore, it is very interesting to investigate
sea-quark shadowing at RHIC. At this stage, the Fermilab Drell-Yan
data are not shown in the shadowing region ($x<0.03$).

Nuclear gluon distributions are also not understood.
The only explicit experimental data are those obtained
by the NMC in muon-induced J/$\psi$ productions, but
accuracy is not good enough to test 10\% effects.
Currently, two-jet events are being analyzed by
the Fermilab muon group for investigating
gluon shadowing. This group may produce interesting results.
In future, proposed RHIC direct-photon experiment should
clarify the gluon-shadowing problem.

$~~~~~$

\hspace{3.0cm}
\begin{tabular}{||l|l|l||} \hline
\end{tabular}

\vspace{0.2cm}

\hspace{3.0cm}
Table 1  Status of nuclear parton distributions.

$~~~~~$

We investigate gluon distributions in the carbon and tin nuclei
by using the $Q^2$ rescaling model with
parton recombination effects.
We obtain strong shadowing in the small $x$ region
due to the recombinations.
The ratio $G_A(x)/G_N(x)$ in the medium $x$ region
is typically 0.9 for medium size nuclei.
At large $x$, the ratio becomes large due to
gluon fusions from different nucleons.

\vspace{-0.46cm}
\begin{list}{}{\leftmargin 0.0in \rightmargin 3.3in}
\item
$~~~~$
Calculated gluon distributions are compared
with the NMC data for $G_{Sn}(x)/$
$G_C(x)$ in Fig. 3.
The dashed curve shows recombination
results with the cutoff for parton leaking
$z_0$=0 and the solid (dash-dot)
curve shows combined results of the $Q^2$ rescaling
and the recombinations with $z_0$=0 ($z_0=2$ fm).
Comparisons with NMC data
for $G_{Sn} (x)/ G_C (x)$
indicate that more accurate experimental data
are needed for testing the model.
\end{list}
\vspace{-1.0cm}
\hspace{8.3cm}
Fig. 3 Nuclear gluon distributions [4].

\vspace{0.3cm}
{\bf [Summary]}
We find that nuclear structure functions $F_2(x,Q^2)$
can be explained in the wide $x$ range
by using a parton model with
$Q^2$-rescaling and parton-recombination effects.
The model is extended to studies of nuclear gluon distributions.
We need detailed theoretical and experimental analyses of nuclear
sea-quark and gluon distributions. Furthermore, valence-quark
distributions at small $x$ could be interesting for studying
nuclear shadowing mechanism.

\vfill\eject
%%%%%%%%%%%%%%%%%%%%%%%%%%%%%%%%%%%%%%%%%%%%%%%%%%%%%%%%%%%%%%%%%%%%%%
%%%%%%%%%%%%%%%%%%%%%%%%%%%%%%%%%%%%%%%%%%%%%%%%%%%%%%%%%%%%%%%%%%%%%%

\noindent
{\bf 2. SU(2)-flavor-symmetry breaking in nuclear antiquark distributions}

\vspace{0.3cm}

Violation of the Gottfried sum rule
was suggested by the NMC
in deep inelastic muon scattering.
It indicates that antiquark distributions in the nucleon
are not SU(2)-flavor symmetric [$\bar u(x)\ne \bar d(x)$].
As an independent test, Drell-Yan data for the tungsten target
have been used for examining the flavor asymmetry.
However, nuclear effects are possibly significant
because the tungsten is a heavy nucleus.
If the nuclear modification is very large, the Drell-Yan analysis
cannot be directly compared with the NMC result.
We discuss the SU(2)-flavor-asymmetric distribution $(\bar u-\bar d)_A$
in nuclei, especially in the tungsten nucleus.

We investigate whether there exists significant modification
of the $\bar u -\bar d$ distribution in nuclei
in a parton recombination model [5].
It should be noted that a finite $\bar u-\bar d$ distribution
is theoretically possible in nuclei
even if the sea is $SU(2)_f$ symmetric in the nucleon.
In neutron-excess nuclei such as the tungsten,
there exist more $d$-valence quarks than $u$-valence quarks, so that
more $\bar d$-quarks are lost than $\bar u$-quarks are due
to parton recombinations in the small $x$ region.
In a parton-recombination picture, partons in different nucleons
could interact in a nucleus.
This is an extra effect which does not exist in a single nucleon.
In discussing antiquark distributions in nuclei, this effect should
be taken into account properly.
Even if the nucleon sea is SU(2) flavor symmetric,
it is interesting to find that a finite flavor asymmetric
distribution can be obtained in a nucleus:
$$
x[\Delta \bar u(x) - \Delta \bar d(x)]_A=
\varepsilon
{ {4K} \over 9 } x \int_0^1 dx_2   ~ x \bar u^* (x)
        ~ x_2 [u_v(x_2)-d_v(x_2)]~  {{x^2+x_2^2} \over {(x+x_2)^4}}
{}~~,
\eqno{(1)}
$$
where $u_v(x)$ and $d_v(x)$ are u and d
valence-quark distributions in the proton,
the neutron-excess parameter $\varepsilon$
is defined by $\varepsilon ={{N-Z} \over {N+Z}}$,
and $K$ is given by $K=9A^{1/3}\alpha_s(Q^2)/$
$(2R_0^2Q^2)$ with $R_0$=1.1 fm.
The physics meaning of the finite distribution is as follows.
In a neutron-excess nucleus ($\varepsilon >0$),
the $d_v$ quark number is larger than the $u_v$ quark one.
Hence, more $\bar d$ quarks are lost than $\bar u$ quarks
in parton recombination processes.
The modification of the flavor-asymmetric distribution
is directly proportional to
the neutron-excess parameter.

We evaluate Eq. (1) with the input
parton distributions MRS-D0 (1993)
in the tungsten nucleus $_{74}^{184} W_{110}$.
The $SU(2)_f$-symmetric-sea distribution in the nucleon
is used by setting $\Delta=0$ in the MRS-D0 distribution.
Obtained results are shown in Fig. 4, where the solid curve shows
the $x[\Delta\bar u-\Delta\bar d]_A$ distribution (per nucleon)
of the tungsten nucleus in Eq. (1).
As expected in a neutron-excess nucleus, the parton recombinations
produce a finite $SU(2)_f$-breaking antiquark distribution even if
they are $SU(2)_f$ symmetric in the nucleon.
Furthermore, it is a positive contribution to $\bar u-\bar d$
because of the d-valence-quark excess over u-valence
in the neutron-excess nucleus.

\vfill\eject
\vspace{-0.46cm}
\begin{list}{}{\leftmargin 0.0in \rightmargin 3.7in}
\item
$~~~~$
The situation is changed if the $SU(2)_f$ asymmetric distribution
is used as the input distribution as shown
by the dashed curve in Fig. 4.
In this case, $\bar q G \rightarrow \bar q$ contributions
become larger than $q \bar q \rightarrow G$ ones,
and obtained results are much different.
The whole contribution becomes
negative at small $x$
due to the $\bar d$-excess over $\bar u$ in the proton.
On the contrary, we find $\bar u$-excess over $\bar d$ in the larger $x$
region ($x>0.05$).
\end{list}
\vspace{-1.0cm}
\hspace{6.6cm}
Fig. 4 Nuclear modification of $\bar u(x)-\bar d(x)$ [5].

\vspace{0.25cm}

{\bf [Summary]}
We find that a finite flavor-breaking distribution
in a nucleus ($[\bar u(x)-\bar d(x)]_A \ne 0$) is possible
even though it is symmetric in the nucleon ($\bar u(x)-\bar d(x)=0$).
Because the Drell-Yan experiments are in progress at Fermilab,
nuclear effects on the flavor asymmetric distribution
could be an interesting topic for future theoretical and
experimental investigations.

%%%%%%%%%%%%%%%%%%%%%%%%%%%%%%%%%%%%%%%%%%%%%%%%%%%%%%%%%%%%%%%%%%%%%%
%%%%%%%%%%%%%%%%%%%%%%%%%%%%%%%%%%%%%%%%%%%%%%%%%%%%%%%%%%%%%%%%%%%%%%
\vspace{0.2cm}

\begin{center}
{\bf Acknowledgment} \\
\end{center}
\vspace{-0.2cm}

S. K. would like to thank the Yamada Science Foundation
for their financial support for his participating in this conference.
This research was partly supported by the Grant-in-Aid for
Scientific Research from the Japanese Ministry of Education,
Science, and Culture under the contract number 06640406.

$~~~$

\noindent
* Email: kumanos@himiko.cc.saga--u.ac.jp.

%%%%%%%%%%%%%%%%%%%%%%%%%%%%%%%%%%%%%%%%%%%%%%%%%%%%%%%%%%%%%%%%%%%%%%
%%%%%%%%%%%%%%%%%%%%%%%%%%%%%%%%%%%%%%%%%%%%%%%%%%%%%%%%%%%%%%%%%%%%%%

\vspace{0.1cm}

\begin{center}
{\bf References} \\
\end{center}

\vspace{0.3cm}

\vspace{-0.30cm}
\vspace{-0.38cm}
\begin{description}{\leftmargin 0.0cm}
\vspace{-0.38cm}
\item{[1]}
S. Kumano and F. E. Close,
               Phys. Rev. {\bf C41}, 1855 (1990);
S. Kumano,in Proceedings of the International Workshop
          on Gross Properties of Nuclei and Nuclear Excitations,
          Hirschegg, Austria, Jan. 20--25, 1992,
                            edited by H. Feldmeier.

\vspace{-0.38cm}
\item{[2]}
S. Kumano, Phys. Rev. {\bf C48}, 2016 (1993) \& {\bf C50}, 1247 (1994).

\vspace{-0.38cm}
\item{[3]}
R. Kobayashi, M. Konuma, and S. Kumano,
preprint SAGA-HE-63-94, submitted for publication;
M. Miyama and S. Kumano, research in progress.

\vspace{-0.38cm}
\item{[4]}
S. Kumano, Phys. Lett. {\bf B298}, 171 (1993).

\vspace{-0.38cm}
\item{[5]}
S. Kumano, preprint SAGA-HE-67-94, submitted for publication.

\end{description}
\end{document}